\documentclass[runningheads]{llncs}

\usepackage{hyperref}
\usepackage[utf8]{inputenc} 
\usepackage[T1]{fontenc}
\usepackage{lmodern}
\usepackage{amsmath}
\usepackage{amssymb}
\usepackage{gensymb}
\usepackage{listings} 
\lstdefinestyle{Muli}{
	language=Java,
	morekeywords={free,fail,solve,getAllSolutions,getAllSolutionsEx,getOneSolution,getOneSolutionEx,muli,search},
	captionpos=b,
	tabsize=2,
	abovecaptionskip={10\p@},
	belowcaptionskip={0\p@},
	showstringspaces=false,
	basicstyle=\ttfamily\footnotesize 
}
\usepackage{cleveref} 
\Crefname{section}{Sect.}{Sections}
\Crefname{listing}{Listing}{Listings}
\Crefname{table}{Tab.}{Tables}
\Crefname{figure}{Fig.}{Figures}
\renewcommand{\figurename}{Figure}
\usepackage{microtype}

\usepackage{cleveref}
\usepackage{graphicx} 
\usepackage{tikz} 
\usetikzlibrary{calc}
\usetikzlibrary{positioning} 
\usetikzlibrary{arrows} 
\usetikzlibrary{shapes.geometric} 
\usetikzlibrary{matrix,fit} 
\usepackage{tikz-qtree}
\usepackage{tikz-uml}
\usepackage{tikzscale} 
\usepackage{erciscolors}

\usepackage{booktabs} 
\usepackage{siunitx}

\usepackage{pgfplots}
\def\addlegendimage{\csname pgfplots@addlegendimage\endcsname}
\pgfkeys{/pgfplots/number in legend/.style={%
		/pgfplots/legend image code/.code={%
			\node [anchor=east]at (0.7,-0.0225){#1}; 
		},%
	},
}

\usepackage{subcaption}

\begin{document}
	\renewcommand{\thelstlisting}{\arabic{lstlisting}} 



	%
	%
        \title{Constraint-Logic Object-Oriented Programming with Free Arrays}


	\author{Jan C. Dageförde\orcidID{0000-0001-9141-7968}
	\and
	Herbert Kuchen
	}
	\authorrunning{J. C. Dageförde and H. Kuchen}
	\institute{ERCIS, Leonardo-Campus 3, 48149 Münster, Germany\\
		\email{\{dagefoerde,kuchen\}@uni-muenster.de}
	}

	%

	\maketitle              

	\begin{abstract}
	Constraint-logic object-oriented programming provides a useful symbiosis between object-oriented programming and constraint-logic search. 
	The ability to use logic variables, constraints, non-deterministic search, and object-oriented programming in an integrated way facilitates the combination of search-related program parts and other business logic in object-oriented applications. 
	With this work we conceptualize array-typed logic variables (``free arrays''), thus completing the set of types that logic variables can assume in constraint-logic object-oriented programming.
	Free arrays exhibit interesting properties, such as indeterminate lengths and non-deterministic accesses to array elements.

	\keywords{constraint-logic object-oriented programming \and free arrays \and non-deterministic element access \and reference types.}
	\end{abstract}

\section{Motivation} \label{sec:intro}

In constraint-logic object-oriented programming (CLOOP), one of the remaining missing puzzle pieces is the ability to use logic variables in lieu of arrays.
As a novel paradigm, CLOOP describes programming languages that add constraint-logic features on top of an object-oriented syntax.
Most importantly, CLOOP offers logic variables, constraints, and encapsulated non-deterministic search, seamlessly integrated with features from object-oriented programming.
As a blueprint for CLOOP languages,
the \textbf{Mu}enster \textbf{L}ogic-\textbf{I}mperative Programming Language (Muli) is a Java-based language that has been successfully used in the generation of artificial neural networks \cite{Dageforde2020applications}, for search problems from the domain of logistics, and for classical search problems \cite{Dageforde2019cola}.
So far, logic variables in Muli can be used instead of variables of primitive types \cite{Dageforde2019cola} or in place of objects \cite{Dagefoerde2020freeobjects}.
Adding support for array-type logic variables is another step on the path to achieving the full potential of CLOOP.
Potential opportunities are illustrated with the code snippet in \Cref{lst:access-nondet}. This snippet declares a logic array \lstinline|a|, i.\,e., an array with an indeterminate number of elements and none of the elements are bound to a specific value.
Moreover, it uses logic variables as indexes for accessing array elements, resulting in non-deterministic accesses.

\begin{lstlisting}[label={lst:access-nondet}, caption={Snippet in which an array as well as the indexes for access are not bound.},float=t]
int i free;
int j free;
int[] a free;
if (a[i] > a[j]) a[i] = a[j] else ...;
\end{lstlisting}

Prior to this work, Muli supported the use of arrays with fixed lengths and logic variables as elements.
In contrast,
free arrays are 
logic variables with an array type that are not bound to  specific values, i.\,e., the entire array is treated symbolically.
In a free array, the individual elements as well as the number of the elements are not known.
This work discusses the introduction of free arrays into CLOOP and Muli.
The paper  starts off by providing a short introduction to the Muli programming language in \Cref{sec:muli}. Afterwards, it presents the contributions of this work:

\begin{itemize}
    \item \Cref{sec:free-arrays} introduces and defines the concept of free arrays in a CLOOP language.
    \item \Cref{sec:free-index} discusses how to handle non-deterministic accesses to array elements when a free variable is used as the index.
    \item These ideas are accompanied by an outline of how free arrays can be implemented in the runtime environment of Muli, specifying the handling of symbolic array expressions as well as the modified behaviour of array-related bytecode instructions (see \Cref{sec:implementation}).
\end{itemize}

\Cref{sec:related-work} presents related work, followed by a short summary in \Cref{sec:conclusion}.

\section{Constraint-logic Object-oriented Programming with Muli} \label{sec:muli}

Our proposal is based on the constraint-logic object-oriented programming language Muli, which facilitates the integrated development of (business) applications that combine deterministic program logic with non-deterministic search. 
Muli is based on Java 8 and adds features that enable constraint-logic search \cite{Dageforde2019cola}.
A key feature of Muli is the ability to declare logic variables. Since logic variables are not \textit{bound} to a specific value, they are called \textit{free variables}. A free variable is declared using the \lstinline|free| keyword, as shown in the following example: \par
    \lstinline|int size free|.\par\hspace{-\parindent}%
Syntactically, declaring a free integer array is valid, too:\par
    \lstinline|int[] numbers free|,\par\hspace{-\parindent}%
however, the behaviour of free arrays is not defined yet, so such a declaration will currently result in an exception at runtime.

Following its declaration, a variable can be used in place of other (regular) variables of a compatible type, for instance as part of a condition: \par
    \lstinline|if (size > 5)|\par\hspace{-\parindent}%
As \lstinline|size| is not bound, the condition can be evaluated to \lstinline|true| as well as to \lstinline|false|, given appropriate circumstances. Upon evaluation of that condition, the executing runtime environment non-deterministically takes a decision and imposes an appropriate constraint that supports and maintains this choice (for example, \lstinline|size > 5| in order to evaluate the \lstinline|true|-branch). 
To that end, the runtime environment leverages a constraint solver for two purposes: First, the constraint solver is queried to check whether the constraint system of an application is consistent, thus avoiding the execution of branches whose constraint system cannot be solved. Second, the constraint solver is used to find specific values for free variables  that respect the imposed constraints.

Eventually, the runtime environment considers all alternative decisions.
The result is a (conceptual) search tree, in which the inner nodes correspond to the points at which decisions can be taken, with one subtree per decision alternative \cite{DagefoerdeTeegen2020}. The eventual outcomes of execution (in particular, returned values and thrown exceptions) are the leaves of the tree.
A returned value or a thrown exception is a solution of non-deterministic search. In addition, Muli provides the facility to explicitly cut execution branches that are not of interest by invoking the \lstinline|Muli.fail()| method. 

The execution behaviour of Muli applications is, for the most part, deterministic and additionally provides \textit{encapsulated search}. Application parts that are intended to perform non-deterministic search need to be declared explicitly in the form of methods or lambda expressions.   These parts are called \textit{search regions}. In order to start search, a search region is passed to an encapsulated search operator (e.\,g., \lstinline|Muli.getAllSolutions()|) that causes the runtime to perform search while collecting all found solutions. After execution finishes, the collected solutions are returned to the invoking (deterministic) part of the application.
Exemplarily, consider the search region presented in \Cref{lst:simple-muli-program}. For a logic variable \lstinline|number| it imposes constraints s.\,t. $0 \leq $ \lstinline|number| $ \leq 5$ by cutting execution branches that do not satisfy this constraint. Otherwise, the symbolic expression \lstinline|number*2| is returned and collected by \lstinline|Muli.getAllSolutions()|, i.\,e., the presented search region returns the numbers $\{0,2,4,6,8,10\}$.

\begin{lstlisting}[label={lst:simple-muli-program}, caption={Search region that imposes constraints on a free variable \lstinline|number| and returns an expression as its solution.},float=t]
Solution<Integer>[] evens = Muli.getAllSolutions(() -> {
	int number free;
	if (number > 5) {
	    throw Muli.fail();
	} else if (number < 0) {
		throw Muli.fail();
	} else {
	    return number*2; } }
\end{lstlisting}

Muli applications are executed on the Münster Logic Virtual Machine (MLVM) \cite{Dageforde2019cola}. The MLVM is a custom Java virtual machine with support for symbolic execution of Java/Muli bytecode and non-deterministic execution of search regions.
The MLVM represents non-deterministic execution in a search tree, in which the inner nodes are \lstinline{Choice} nodes (with one subtree per alternative decision that can be taken) and the leaf nodes are outcomes of search, i.\,e., solutions or failures \cite{DagefoerdeTeegen2020}.
Executing a bytecode instruction with non-deterministic behaviour results in the creation of a \lstinline{Choice} node that is added to the search tree. For example, executing an \lstinline{If_icmpeq} instruction (that corresponds to evaluating an equality expression as part of an \lstinline{if} condition) results in the creation of a  \lstinline{Choice} node with two subtrees, one per alternative outcome, provided that the result of \lstinline{If_icmpeq} can take either value according to the constraints that have already been imposed.

\section{Arrays as Logic Variables} \label{sec:free-arrays}

Muli relies on the symbolic execution of Java/Muli bytecode, i.\,e., symbolic expressions are generated during the evaluation of expressions that cannot (yet) be evaluated to a single constant value. 
Therefore, adding support for free arrays implies introducing symbolic arrays into the execution core of the MLVM.

The length of arrays in Java (and, therefore, in Muli) does not need to be known at compile time, as the legal code example in \Cref{lst:length-at-compile-time} demonstrates: The number of elements that the array \lstinline|arr| holds will become known at runtime.
The length is arbitrary, provided that it can be represented by a positive (signed) \lstinline{int} value \cite{jvms8}.\footnote{At least in theory, as the \lstinline{Newarray} bytecode instruction takes an \lstinline{int} value for the length. In practice, the actual maximum number of elements may be lower as it depends on  the available heap size on the executing machine.}
As a consequence, a free array that is declared using \par%
\lstinline{T[] arr free}\par\hspace{-\parindent}%
comprises
\begin{itemize}
    \item an unknown number of elements, so that  \lstinline{arr.length} is a free \lstinline{int} variable $n$, where $0 \leq n \leq $ \lstinline{Integer.MAX_VALUE}, and
    \item one free variable of type \lstinline{T} per element \lstinline{arr[i]} with \lstinline{i} $<$ \lstinline{arr.length}
\end{itemize}

\begin{lstlisting}[label={lst:length-at-compile-time},caption={The length of an array is not necessarily known at compile time. This example snippet determines the length at runtime instead.},float=t]
ArrayList<Integer> list = new ArrayList<>();
for (int i = 0; i < (int)(Math.random()*1000); i++)
	list.add(i);
int[] arr = new int[list.size()];
\end{lstlisting}

Treating the length of a free array \lstinline{arr} as a free variable provides the benefit that the length can be influenced by imposing constraints  over \lstinline{arr.length}, i.\,e., by referring to the length as part of \lstinline|if| conditions.
Moreover, for an array 
\lstinline|T[] arr free|
the type of the individual array elements \lstinline{arr[i]} depends on what \lstinline{T} is:
    \begin{itemize}
        \item If \lstinline{T} is a primitive type, each element is a simple free variable of that type.
        \item If \lstinline{T} is an array type, the definition becomes recursive as each element is, in turn, a free array.
        \item If \lstinline{T} is a class or interface type, each element is a free object. Therefore, the actual type \lstinline{T'} of each element is \lstinline{T'} $ \preceq $ \lstinline{T}, i.\,e., an element's type is either \lstinline{T} or a type that extends or implements \lstinline{T}. We do not go into specifics on free objects, as they are not of particular relevance here. The interested reader is directed to \cite{Dagefoerde2020freeobjects}  on that matter.
    \end{itemize}

Java requires regular arrays to be initialized 
  either using an array creation expression of the form \lstinline{T[] arr = new T[n];}, resulting in an array \lstinline{arr} with \lstinline{n} elements of type \lstinline{T} \cite[\S\ 15.10.1]{jls8};
  or an array initializer such as \lstinline|int[] arr =| \allowbreak\lstinline|{1, 2};|, resulting in an integer array that holds exactly the specified elements \cite[\S\ 10.6]{jls8}.
For free arrays, this opens up alternative ways of declaring (and initializing) a free array in Muli.

\begin{description}
\item[Simple free variable declaration]
First, following the syntax that is used to declare any free variable,
\lstinline{T[] arr free} declares a free array whose length and elements are indeterminate.

\item[Modified array creation expression]
Second, \lstinline{T[] arr = new T[n] free;} is a modified array creation expression that allows to specify a fixed length for the array (unless \lstinline{n} is free) while refraining from defining any of the array elements.

\item[Modified array initializer]
Third, a modification of the array initializer expression facilitates specifying the length as well as some array elements that shall be free; e.\,g., \lstinline|int[] a = {1, free, 0};| would define an array with a fixed length with two constant elements and a free one at \lstinline{a[1]}.
Trivially, regardless of the chosen initializer, array elements can be modified after the array has been initialized using explicit assignment. For example, 
\lstinline|a[1] = 2;| can be used to replace an element (for example, a free variable) with a constant, and \lstinline|int i free; a[1] = i;| replaces the element at index 1 with a free \lstinline{int} variable.
\end{description}

These considerations facilitate the initialization and subsequent use of logic variables that represent arrays or array elements.
All three alternatives are useful and should therefore be syntactically valid. 
For example, 
\Cref{lst:init-and-constraint} combines the initialization of a free string array via a simple free variable declaration, followed by  imposing a constraint over the array's length (with \lstinline|Muli.fail()|  effectively cutting the branch of execution in which that constraint would not be satisfied).

\begin{lstlisting}[label={lst:init-and-constraint}, caption={Limiting a free array's length to at most five elements by imposing an appropriate constraint.},float=t]
String[] outputLines free;
if (outputLines.length > 5) {
    throw Muli.fail();
} else {
    // <...>
}
\end{lstlisting}

\section{Non-deterministic Access to Array Elements} \label{sec:free-index}

Reconsider the example snippet from a search region that is given in \Cref{lst:access-nondet}: Free arrays become particularly interesting when array elements are accessed without specifying the exact index, i.\,e., with the index as a free variable (e.\,g., \lstinline{arr[i]} where \lstinline{int i free}).
In the comparison \lstinline|a[i] > a[j]|, the array \lstinline|a| as well as the indexes for access are free variables.
For a more complex example, consider the application depicted in  \Cref{lst:simplesort}. It shows a simple sorting algorithm. The algorithm is not particularly efficient, but rather serves to show how free arrays can be used in a Muli application and demonstrates the use of other Muli features as well.
The general idea of \Cref{lst:simplesort} is to find a permutation of the elements of \lstinline|b| that leads to a sorted array \lstinline|a|. 
In line 4 of \Cref{lst:simplesort}, a free array of indexes is introduced. In lines 9--11, the unbound elements of this array are used as indexes of the arrays \lstinline{usedIdx} and \lstinline{a}. The algorithm searches for a permutation s.\,t. the final array is sorted. Consequently, if an index is used more than once, the array \lstinline{idx} does not represent a  permutation and the current branch of the search fails (line 9). Then, another branch is tried after backtracking. If the considered permutation does not lead to a sorted array, the current branch of the search also fails, thus resulting in backtracking (line 13). The efficiency of the algorithm ultimately depends on the constraint solver on which the Muli runtime system relies. Currently, Muli offers using either JaCoP \cite{jacop:Kuchcinski2003} or a custom SMT solver from the Münster Generator for Glass-box Test Cases (Muggl \cite{Ernsting2012}). Moreover, the MLVM provides a flexible solver component that facilitates the addition of alternative 
constraint solvers to the MLVM \cite{Dageforde2018mulisac}.

\begin{lstlisting}[label={lst:simplesort}, caption={Simple sorting algorithm that leverages free arrays.}, numbers=left,float=t]
class SimpleSort {
  public static double[] sort(double[] b) {
    int n = b.length;
    int[] idx free;
    boolean[] usedIdx = new boolean[n];
    for (int i=0; i < n; i++) usedIdx[i] = false;
    double[] a = new double[n];
    for (int i=0; i < n; i++) {
      if (usedIdx[idx[i]]) throw Muli.fail();
      a[idx[i]] = b[i];
      usedIdx[idx[i]] = true; }
    for (int i=0; i < n-1; i++)
      if (a[i] > a[i+1]) throw Muli.fail();
    return a; }

  public static void main(String[] args) {
    double[] b = {42.0, 17.0, 56.3, 78.1, 5.9, 27.2};
    Solution<double[]> a = Muli.getOneSolution( () -> sort(b) );
    for (int i=0; i< a.value.length; i++) {
      System.out.print(a.value[i]); } } }
\end{lstlisting}

Accessing an array with a free index is a non-deterministic operation, because more than one array element (or even a thrown runtime exception) could be the result of the access operation.
Subsequently,
we present approaches that can be used for handling such non-deterministic accesses to arrays. This list of approaches is probably non-exhaustive as there may be additional alternatives.

A first and simple approach would be to branch over all possible values for the index \lstinline{i} in case that there is an access to \lstinline{a[i]} where \lstinline|i free|.
Effectively, this is the equivalent of a labeling operation, 
successively considering every array element as the result of the access.
 Clearly,
this would lead to a huge search space and it is hence not a reasonable option in most cases.

A second approach could store constraints that involve accesses to array elements with unbound indexes symbolically.
In the example from \Cref{lst:access-nondet}, this implies storing the expression \lstinline{a[i] < a[j]} as a constraint.
This approach is complex to handle.
In our example, it would require that, after every change in the remaining domains for \lstinline{i} or \lstinline{j}, we would have
to check whether there are still possible values for \lstinline{i} or \lstinline{j} such that \lstinline{a[i] < a[j]} can be satisfied.
In the worst case
that means that we have to check the constraint for all remaining pairs of values for \lstinline{i} and \lstinline{j}. As a consequence, this approach would be nearly as
complex as the first one, the only difference being that the satisfiability check can stop as soon as values for \lstinline{i} or \lstinline{j}
have been detected which satisfy the constraint. 

A third approach could delay the check of constraints with symbolic array expressions until the involved indexes assume concrete, constant values.
This would be similar to the delayed processing of negation in Prolog \cite{AB94}.
However, in contrast to Prolog,
the ongoing computation would still continue.
At the latest, the constraint needs to be checked when leaving the encapsulated
search space, 
possibly after implicit labeling.
Alternatively, the Muli application could explicitly demand checking delayed constraints, and the MLVM could throw an exception
indicating that there are still delayed constraints when trying to leave an encapsulated search before a check.
This approach is relatively easy to integrate into the MLVM.
However, a major disadvantage of this approach is that
time is wasted for exploring parts of the search space which could have been excluded if we had checked the constraint earlier (and found it to be unsatisfiable).
Even worse, the corresponding computations could have caused external side-effects which should have never happened.
This is a problem since external side-effects cannot be reverted on backtracking (e.\,g., file accesses or console output).
Hence, they are discouraged in encapsulated search regions, especially in the case of delayed constraints. 
Moreover, there is no guarantee that checking the constraint at the end is easier than checking it immediately:
If no additional constraints over \lstinline{i} and \lstinline{j} are encountered in further evaluation,  \lstinline{i} and \lstinline{j} may still assume the same values. Therefore, the delayed evaluation of the initial constraint is just as complicated as a strict evaluation.

A fourth and last approach could entirely forbid constraint expressions that involve unbound variables as array indexes. However, we feel that this
approach is too restrictive. Moreover, it would not really provide new possibilities in Muli.

Unfortunately, all approaches that we could think of have some disadvantages. After comparing the advantages and
disadvantages, we plan to implement the second and third approach which seem most suitable to us. 
As a consequence, the Muli runtime is going to be able to allow developers to configure the system 
in order to choose an approach that best suits their respective search problem.
A quantitative evaluation will be able to show whether one approach is generally favourable over the other.

\section{Implementation} \label{sec:implementation}

Implementing the above considerations affects two areas of the runtime environment:
First, the solver component must be capable of dealing with constraints over free arrays, i.\,e., it must be able to check a constraint system that comprises such constraints for consistency as well as to find values for the involved variables. 
Second, the execution core requires a modified execution semantics of array-related bytecode instructions.
Subsequently, we outline a concept for an implementation in the MLVM.

\subsection{Modelling Constraints over Free Arrays}

\begin{figure}[t]
	\centering
	\includegraphics{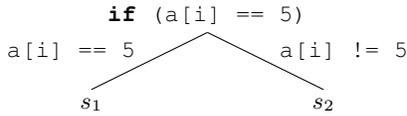}
	\caption{Excerpt from a search tree, showing branch constraints that involve a symbolic expression for array element access, namely, \lstinline|a[i]|.}
	\label{fig:symbarray-branch}
\end{figure}

Accessing an array element using a free variable as an index,
e.\,g. \lstinline|a[i]| with \lstinline|i free|,
would yield a symbolic array expression (as described in \Cref{sec:bytecode-modifications}).
Using that as part of a condition, e.\,g., \lstinline|if (a[i] == 5) {| $s_1$ \lstinline|} else {| $s_2$ \lstinline|}| causes the runtime environment to branch, thus creating a choice with two branches and appropriate constraints as illustrated in \Cref{fig:symbarray-branch}.

\begin{figure}[t]
	\resizebox{\linewidth}{!}{%
	\includegraphics{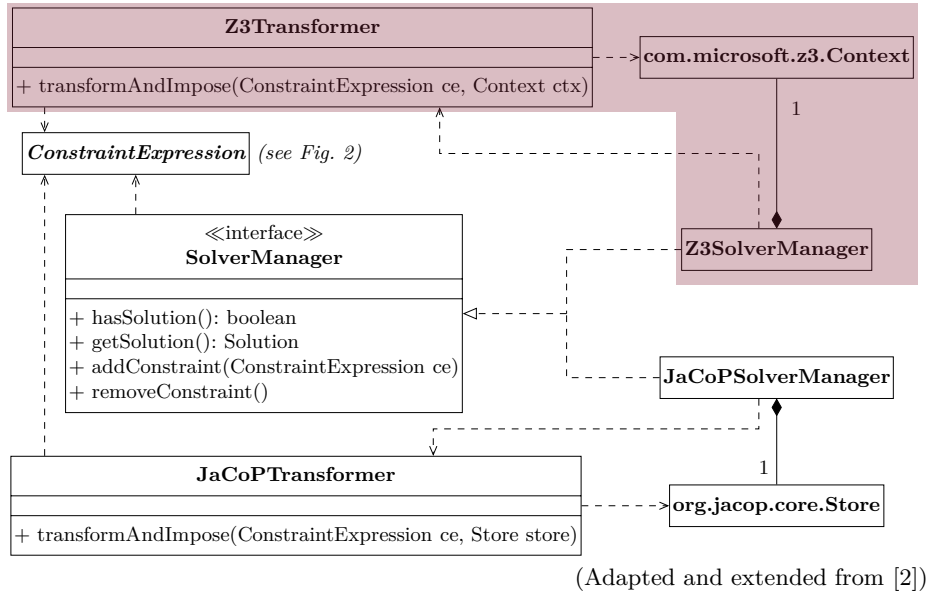}
	}

	\hfill (Adapted and extended from \cite{Dageforde2020diss})

	\caption{Required modification to the solver component of the MLVM in order to integrate the Z3 solver (additions shaded in red).}
	\label{fig:z3-solver}
\end{figure}

The way that a constraint involving symbolic array expressions (such as \lstinline{a[i] <} \lstinline{a[j]} from \Cref{lst:access-nondet})  is modelled depends on the constraint solver.
The two solvers that are currently available in the MLVM do not provide native support for array theories (cf. \cite{MB09}), therefore the MLVM requires additional effort in order to emulate support for such constraints.
Alternatively, the MLVM can leverage a solver that features native support.
For instance, native support is featured by the Z3 solver \cite{MB08,MB09}, which can be used as an incremental constraint solver.\footnote{This is helpful as constraints are added (and removed) incrementally during encapsulated search.}
For handling constraints, the MLVM implements a \textit{solver component} that abstracts from the actual underlying solver.
This is achieved by offering a unified interface for the definition of symbolic expressions and constraints.
Using a set of transformation methods, the defined constraints are transformed into a suitable representation for the respective solver that can then be queried from the MLVM using an adapter-pattern implementation.
As a consequence, it is possible to add support for symbolic array expressions to the unified interface as illustrated in \Cref{fig:symbolic-array-expression}.
Based on that, we can then proceed with both alternatives, i.\,e., implement support for using Z3 as the solver (see \Cref{fig:z3-solver}), but also add transformation routines for symbolic array expressions to the existing solver adapters.

If a solver is used that does not natively provide support for array theories, 
we need to implement a check whether there is at least one binding for the involved index variables for which a constraint involving a symbolic array expression holds. This check is implemented in the constraint transformation method for a constraint that involves symbolic array expressions.
A generic approach iterates over all index variables $X \in Indexes$, and substitute $X$ for an allowed value from the domain of $X$.
For example, for the symbolic array expression \lstinline|a[i] > a[j]| this would result in the double \lstinline|for| loop presented in \Cref{lst:workaround} that iterates over all possible bindings for \lstinline|i| and \lstinline|j| and checks a simplified constraint with a constant index together with the active constraint system.
If a binding is encountered that satisfies the constraint system, the check returns \lstinline|true| because a single binding suffices.
Otherwise, if no such binding is encountered, the check returns \lstinline|false| to indicate that the constraint system is not satisfiable.

\begin{lstlisting}[language=,caption={Checking a symbolic array expression constraint by checking simplified constraints that involve only concrete array elements.},label=lst:workaround,float=t]
for(x in Domain(i)) {
	for (y in Domain(j)) {
		store.impose(a[x] > a[y]);
		if (store.isConsistent()) return true;
		else store.remove(a[x] > a[y]);	} }
// No binding satisfies the constraint system.
return false;
\end{lstlisting}

In order to support the third approach from \Cref{sec:free-index}, i.\,e. delayed constraint checking, the above constraint transformations would only be performed as soon as labelling is required or as soon as one of the index variables has a singleton domain, making the index effectively constant.

\begin{figure}[t]
	\resizebox{\linewidth}{!}{%
	\includegraphics{fig/symbolic-array-expression.tikz}
	}

	\hfill (Adapted and extended from \cite{Dageforde2020diss})
	
	\caption{Augmenting the unified interface for the definition of constraints and expressions in order to add symbolic array expressions (additions shaded in red).}
	\label{fig:symbolic-array-expression}
\end{figure}

As an alternative to implementing the proposed emulation for checking constraints that involve symbolic array expressions, we can integrate the Z3 solver into the MLVM solver component in order to leverage its native support for such constraints.
The required modifications to the solver component are illustrated in \Cref{fig:z3-solver}, by implementing adapter classes and transformation classes that are similar in structure to the way how the JaCoP solver is integrated into the MLVM.

The \lstinline|Context| type is the main class provided by the official Z3 Java bindings \cite{Z3java-context}.  In order to use it from the MLVM, the \lstinline|Z3SolverManager| type serves as an adapter class, implementing the interface that is expected from an MLVM solver manager and delegating calls to an instance of the \lstinline|Context| type appropriately.
The Z3 context instance needs to be configured to use incremental solving in order to properly handle the incremental addition and removal of constraints during encapsulated search.
Moreover, the \lstinline|Z3SolverManager| relies on the \lstinline|Z3Transformer|
in order to transform expressions and constraints specified in the unified interface to a corresponding representation for the Z3 Java bindings.
For instance, the Z3 transformer would transform a symbolic array expression of the form \lstinline|a[i] == y| (where \lstinline|a| is a free integer array and \lstinline|i, y| are free integers) into the following commands for the Z3 solver: 
\begin{lstlisting}[language=lisp,xleftmargin=\parindent]
(declare-const a (Array Int Int))
(declare-const i Int)
(declare-const y Int)
(assert (= y (select a i)))
\end{lstlisting}

The Z3 solver has been successfully used in the context of glass-box test case generation, e.\,g. with the Pex tool \cite{TH08}, even of applications that use symbolic arrays and indices. Therefore, we assume that it will provide adequate performance (and perhaps a better performance compared to our emulation for other solvers).
Nevertheless, implementing all the above alternatives will facilitate an evaluation of their performance.

\subsection{Modifications to Bytecode Execution Semantics}
\label{sec:bytecode-modifications}

The MLVM executes Java bytecode. Implementing the above considerations requires modifications to the execution semantics of the following bytecode instructions:
\lstinline{Newarray}, 
\lstinline{Arraylength}, \lstinline{Xaload}, \lstinline{Xastore} (where \lstinline{X} is replaced with a type, e.\,g., \lstinline{Iastore} to store an array element of type \lstinline{int} \cite{jvms8}.

\lstinline|Newarray| is typically used in order to create an array on the heap. For the case of a free array,
this requires the creation of an internal representation of the free array, 
comprising a \lstinline{NumericVariable} for the \lstinline|length| attribute (so that the length of a free array can become part of symbolic expressions) 
as well as an 
\lstinline{ArrayList<T>} that will hold the individual elements.
This representation will only be used internally by the MLVM. For Muli applications that use a free array its type is equivalent to that of a corresponding regular array.

The \lstinline{Arraylength} bytecode instruction returns the length of an array  \cite[\S~6.5]{jvms8}. If it is executed in the context of a free array, the instruction has to yield the symbolic representation of the free array's length.
As an exception to that, if the logic variable for the length is already bound to a single value, \lstinline|Arraylength| can return a constant.

The modifications to the \lstinline{Xaload} and \lstinline{Xastore} instructions 
work identically regardless of their type \lstinline|X|
and result in (potentially) non-deterministic execution.
The \lstinline{Xaload} instruction is the bytecode equivalent of accessing a single array element, e.\,g., \lstinline|a[i]|, whereas \lstinline{Xaload} is the equivalent of assigning a value to an array element, e.\,g., \lstinline|a[i] = x|.
Execution requires to make a distinction based on what is known about the length $n$ of the involved free array (e.\,g., from constraints that have already been imposed on $n$).
For \lstinline|a[i]|, if \lstinline|i| is definitely within the range $(0..n-1)$, the behaviour is deterministic and returns a symbolic array access expression accordingly.
Similarly, if \lstinline|i| is outside that range, 
the execution (deterministically) results in throwing a runtime exception of the type \lstinline{ArrayIndexOutOfBoundsException}.
In all other cases, execution results in the creation of a  non-deterministic choice, distinguishing successful access (yielding a symbolic expression) and the error case (yielding an exception) as alternative outcomes. Each alternative results in imposing appropriate constraints over \lstinline|i| and $n$.
Using backtracking, the MLVM will evaluate both alternatives successively.

\section{Related Work} \label{sec:related-work}

A first approach to a symbolic treatment of arrays dates back to
McCarthy's \emph{basic theory of arrays} developed  in 1962 \cite{McCarthy62}. It consists of just two axioms, one telling that if a value \lstinline{v} is assigned to \lstinline{a[i]} then \lstinline{a[i]} later on has this value \lstinline{v}. The other axiom essentially says that changing \lstinline{a[i]} does not affect any other array element. These axioms are clearly not enough
for handling free arrays in Muli. McCarthy's approach was extended to the \emph{combinatorial array logic} by de Moura and Bj{\o}rner \cite{MB09}. It is expressed by a couple of inference rules, which work on a more
abstract level and do not address the processing of the search space. 
Nevertheless, these rules are among the theoretical foundations of Microsoft's Z3 SMT solver \cite{MB08}. Based on this solver, support for 
arrays was included into Microsoft's test-case generator Pex \cite{TH08} and
into the symbolic code execution mechanism of NASA's Java Pathfinder, a model checker and test-case generator for Java programs \cite{Fromherz2017}. In order to achieve the latter, Fromherz et al. 
mainly changed the semantics and treatment of the \lstinline|Xaload| and \lstinline|Xastore| bytecode instructions of their symbolic variant of the Java virtual machine. Their changes to these instructions are similar to our intended modifications of the MLVM, with the exception that the MLVM has a more sophisticated mechanism for backtracking and resuming an encapsulated search. The authors do not discuss approaches for dealing with
the potentially huge search space caused by array constraints.

Also in the context of test-data generation, Korel
\cite{Korel1990} presented an array-handling approach which avoids the difficulties of free arrays and symbolic array indexes by resorting to a non-symbolic execution. Korel used a concrete evaluation in combination with dataflow analysis and so-called function minimization in order to reduce the search space.
This approach is not suitable for a CLOOP language.

All the mentioned approaches stem from the domains of test-case generation and model checking. To the best of our knowledge, there is no programming language yet that offers free arrays with symbolic array indexes.

\section{Conclusion and Outlook} \label{sec:conclusion}

As a research-in-progress paper, this work presents approaches for the addition of free arrays to constraint-logic object-oriented programming, thus  starting a discussion that will eventually result in a prototypical implementation for the Muli programming language.
The present paper discusses the characteristics and implementation aspects of free arrays.
In particular, we address the symbolic treatment of the array length and symbolic array indexes based on constraints.
Moreover, we propose a syntax for the declaration and initialization of free arrays.
In addition, we discuss ways of dealing with non-deterministic accesses to array elements, proposing possible solutions to that end.
The proposed concepts facilitate the use of logic arrays in the context of encapsulated, non-deterministic search that is interleaved with deterministic computations. 
Moreover, Muli allows using arbitrary search strategies in order to use symbolic computations that involve arrays.

Future work will implement support for free arrays into the MLVM  based on the approaches presented here.
Moreover, the Z3 solver will be added to the MLVM as an alternative backend of the solver component so that its support for symbolic array expressions can be leveraged.
This allows for an exhaustive evaluation of the approaches in combination with different solvers.

%
%
%
\bibliographystyle{splncs04}
\bibliography{lit}

\end{document}